\documentclass[aps,prl,twocolumn,amsmath,amssymb,superscriptaddress,nofootinbib,longbibliography]{revtex4}
\usepackage{graphicx}
\usepackage{bm}
\usepackage{amssymb}
\usepackage{amsmath}
\usepackage{latexsym}
\usepackage{color}
\usepackage[normalem]{ulem}
\usepackage{cancel}
\newcommand{\vc}[1]{\mbox{\boldmath $#1$}} 
\newcommand{\ind}[1]{_{#1}}    
\newcommand{\indrm}[1]{_{\mathrm {#1}}}    
\newcommand{\sca}{$^{45}$Sc}    
\newcommand{\fem}{$^{57}$Fe}    
\newcommand{\zn}{$^{67}$Zn}    
\newcommand{\ag}{$^{109}$Ag}    
\newcommand{\thr}{$^{229}$Th}    
\newcommand{\sco}{Sc$_{\ind{2}}$O$_{\ind{3}}$}    
\newcommand{\scf}{ScF$_{\ind{3}}$}    
\newcommand{\sam}{ScAlMgO$_{\ind{4}}$}    
\newcommand{\taux}{\tau_{\ind{0}}^{*}}  

\definecolor{highlt}{rgb}{0,0,0.9}
\definecolor{darkblue}{rgb}{0.0,0,0.5}
\definecolor{prublue}{rgb}{0.0,0.49,0.83}
\definecolor{darkgreen}{rgb}{0,0.6,0}

\begin{document}  
\title{Probing the Linewidth of the 12.4-keV $^{45}$Sc Isomeric Resonance in Solids by Nuclear Forward Scattering}

\author{Peifan Liu}
\affiliation{Argonne National Laboratory, Lemont, Illinois, USA}
\author{Miriam Gerharz}
\affiliation{Max Planck Institute for Nuclear Physics (MPIK), Heidelberg, Germany}
\author{Berit Marx-Glowna} 
\affiliation{Helmholtz-Institut Jena, Jena, Germany}
\affiliation{GSI Helmholtzzentrum f\"ur Schwerionenforschung, Darmstadt, Germany}
\author{Willi Hippler}
\affiliation{Friedrich-Schiller-Universit\"at Jena, Jena, Germany}
\author{Jan-Etienne Pudell}
\author{Alexey Zozulya}
\affiliation{European X-Ray Free-Electron Laser Facility, Schenefeld, Germany}
\author{Brandon Stone}
\affiliation{Argonne National Laboratory, Lemont, Illinois, USA}
\author{Deming Shu}
\affiliation{Argonne National Laboratory, Lemont, Illinois, USA}
\author{Robert Loetzsch}
\affiliation{Friedrich-Schiller-Universit\"at Jena, Jena, Germany}
\author{Sakshath Sadashivaiah}
\affiliation{Helmholtz-Institut Jena, Jena, Germany}
\affiliation{GSI Helmholtzzentrum f\"ur Schwerionenforschung, Darmstadt, Germany}
\author{Lars Bocklage}
\author{Christina Boemer}
\author{Shan Liu}
\author{Vitaly Kocharyan} 
\author{Dietrich Krebs}
\author{Tianyun Long}
\author{Weilun Qin}
\author{Matthias Scholz}
\author{Kai Schlage}
\author{Ilya Sergeev}
\author{Hans-Christian Wille}
\affiliation{Deutsches Elektronen-Synchrotron DESY, Hamburg, Germany}
\author{Ulrike Boesenberg}
\author{Gianluca Aldo Geloni}
\author{J\"org Hallmann}
\author{Wonhyuk Jo}
\author{Naresh Kujala}
\author{Anders Madsen}
\author{Angel Rodriguez-Fernandez}
\author{Rustam Rysov} 
\author{Kelin Tasca}
\affiliation{European X-Ray Free-Electron Laser Facility, Schenefeld, Germany}
\author{Tomasz Kolodziej}
\affiliation{National Synchrotron Radiation Centre SOLARIS, Krakow, Poland}
\author{Xiwen Zhang}
\affiliation{Texas A\&M University, College Station, Texas, USA}
\author{Niclas Wieland}
\author{G\"unter Huber}
\affiliation{University of Hamburg, Hamburg, Germany}
\author{James H. Edgar}
\affiliation{Kansas State University, Manhattan, Kansas, USA}
\author{J\"org Evers}
\affiliation{Max Planck Institute for Nuclear Physics (MPIK), Heidelberg, Germany}
\author{Olga Kocharovskaya}
\affiliation{Texas A\&M University, College Station, Texas, USA}
\author{Ralf R\"ohlsberger}
\affiliation{Helmholtz-Institut Jena, Jena, Germany}
\affiliation{GSI Helmholtzzentrum f\"ur Schwerionenforschung, Darmstadt, Germany}
\affiliation{Friedrich-Schiller-Universit\"at Jena, Jena, Germany}
\affiliation{Deutsches Elektronen-Synchrotron DESY, Hamburg, Germany}
\author{Yuri Shvyd'ko} \thanks{Corresponding author: shvydko@anl.gov} 
\affiliation{Argonne National Laboratory, Lemont, Illinois, USA}

\begin{abstract} 
  The $^{45}$Sc transition from the nuclear ground state to the
  12.389-keV isomer (lifetime 0.46~s) has an ultranarrow natural
  linewidth $\Gamma_{\ind{0}}\simeq 1.4$~feV, corresponding to a
  quality factor $\simeq 10^{19}$. This makes $^{45}$Sc a compelling
  candidate for nuclear-clock metrology, provided that coherence in
    solids can be maintained close to the natural limit.  Here we
    investigate the linewidth and coherence properties of the
  $^{45}$Sc resonance in solids following resonant x-ray pumping.
  
  \noindent Using the European XFEL, we confirm persistence of the
  long-lived isomer excitation in a solid-state environment via
  time-delayed incoherent $K_{\alpha,\beta}$ emission and observe a
  weak delayed elastic channel at 12.4~keV, from which we extract a
  partial internal-conversion coefficient
  $\alpha_{K}=390(60)$. Time-domain nuclear forward scattering
  measurements in crystals of Sc, Sc$_2$O$_3$, ScN and ScAlMgO$_4$ at
  20~K show no statistically significant coherent forward-scattering
  signal beyond $2$~ms; within the adopted linewidth-broadening model,
  we infer an effective broadening
  $\Delta\Gamma \gtrsim 500\,\Gamma_{\ind{0}}$.  These results provide
  the quantitative constraints on linewidth broadening in solid-state
  $^{45}$Sc with an intrinsic natural linewidth in the
  femto-electronvolt range thereby laying the groundwork for precision
  metrology in the X-ray regime and future nuclear-clock frequency
  references using \sca\ nuclear isomer.

\end{abstract}

\maketitle

{\em Introduction--}
Oscillators with sharp resonance frequencies and
large quality factors—reference oscillators or frequency standards—are
fundamental to our ability to measure time with precision.
{ Atomic clocks define the unit of time (the second),
  enable GPS functionality, and allow tests of fundamental principles
  of physics \cite{Ludlow15,SBD18}. Atomic optical clocks are currently the
  most precise realization of atomic frequency standards, achieving
  fractional uncertainties as small as $\simeq 10^{-18}$
  \cite{Ludlow15}. They are expected to form the basis of a future
  redefinition of the second.}  The search for more accurate, stable,
and convenient reference oscillators is ongoing.

Nuclear, alongside atomic, resonances are promising candidates for
reference oscillators \cite{BSS21,PSS21}. Nuclear resonances offer
higher transition energies $E_{\ind{0}}$, providing enhanced stability
for clock applications due to statistical advantages. Many nuclear
isomers exhibit extremely long lifetimes $\tau_{\ind{0}} \gtrsim 1$~s,
minuscule natural widths $\Gamma_{\ind{0}}=\hbar /\tau_{\ind{0}}$, and
very large quality factors
$Q_{\ind{0}}=E_{\ind{0}}/\Gamma_{\ind{0}}$. Table~\ref{tab1}
highlights selected examples. Nuclei are also less sensitive to
ambient electromagnetic (EM) fields than atomic electrons because of
their small size, small EM moments, and{, in many
  cases, the partial screening of external fields by the surrounding
  electrons}.  This reduced sensitivity could minimize inhomogeneous
resonance broadening and eliminate the need for isolating clock atoms
in dilute gases or ion traps at low temperatures -- potentially
enabling the use of macroscopic quantities of atoms in solids.

The M\"ossbauer effect enables nuclei in solids to exhibit narrow
spectral resonance lines at moderately low temperatures
\cite{Wertheim64}, offering a promising path toward solid-state
nuclear clocks. While nuclear resonances are relatively insensitive to
external perturbations, they are not immune. In solids, the actual
linewidth $\Gamma=\Gamma_{\ind{0}}+\Delta\Gamma$ can be significantly
broadened ($\Delta\Gamma \gg \Gamma_{\ind{0}}$) by inhomogeneities
such as variations of hyperfine parameters, reducing the actual
quality factor $Q = E_{\ind{0}} / \Gamma$.

Among optical- or higher-frequency resonances { with
resonant-center number densities of order $\sim 10^{23}$~cm$^{-3}$},
the narrowest measured linewidth is $\Gamma=50$~peV (12~kHz) for the
93-keV nuclear transition in \zn\ \cite{PSS92}, matching its natural
width $\Gamma_{\ind{0}}$.
This is two orders of magnitude
narrower than the widely used 14.4-keV resonance of \fem\
\cite{Wertheim64}. The 88-keV solid-state resonance in \ag\ has an
indirectly estimated width of $\Gamma \simeq 0.1$~feV (30~mHz)
\cite{BDI09}, only $\simeq 10$ times broader than its natural width
$\Gamma_{\ind{0}}$ and $\simeq 10^{6}$ times narrower than that of \zn
.

The isomeric transition in \thr, with its anomalously low transition
energy of 8.4~eV \cite{WSL16,Thfreq20,SWB19,KMA23}, is considered one
of the most promising nuclear clock reference oscillators. Recent
advancements include laser excitation of the \thr\ nuclear isomeric
transition in a solid-state host \cite{TOZ24,ESJ24} and the
establishment of a frequency link between nuclear and electronic
transitions using a VUV comb to directly measure the frequency ratio
of the \thr\ nuclear clock transition and the $^{87}$Sr atomic clock
with a very small uncertainty of $\simeq 10^{-13}$
\cite{ZOH24}. However, the smallest measured resonance width
$\Gamma \simeq 25$~kHz \cite{ODZ25} is $\simeq 10^{8}$ times broader
than $\Gamma_{\ind{0}}=0.24$~mHz \cite{ZOH24}, reducing dramatically
the quality factor $Q$ compared to its natural value.


\begin{figure*}[t!]
\includegraphics[width=1.0\textwidth]{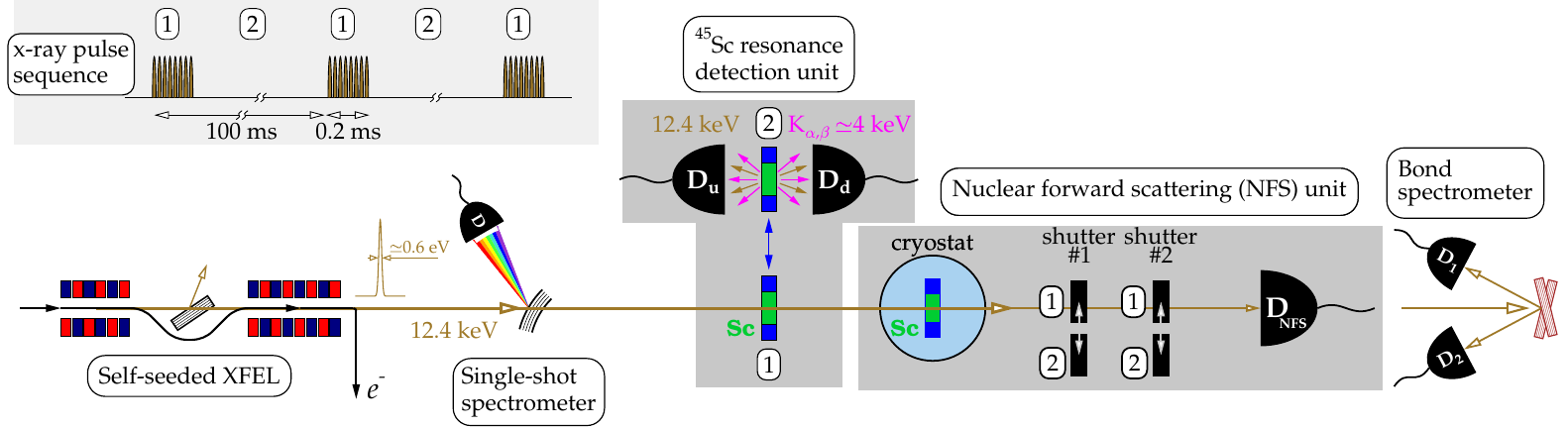}
\caption{Schematic of the experiment, designed to resonantly excite
  \sca\ nuclei with 12.4-keV x-ray pulses and to detect both
  coherently and incoherently scattered time-delayed photons. X-ray
  macropulses from the self-seeded XFEL sequentially excite \sca\
  nuclei in two targets: (i) a metallic Sc foil that shuttles within
  the \sca\ resonance-detection unit between the x-ray beam and the
  detectors D${\indrm{u}}$ and D${\indrm{d}}$, and (ii) a crystal
  target in the nuclear forward scattering (NFS) unit, cooled by a
  cryostat to 20~K. Nuclear decay products—12.4-keV photons and Sc
  $K_{\ind{\alpha,\beta}}$ fluorescence photons—are measured by x-ray
  detectors D${\indrm{u}}$, D${\indrm{d}}$, and D$_{\indrm{NFS}}$ as a
  function of the re-emitted photon energy $E$ and the time delay $t$
  after excitation (see Figs.\ref{fig2} and \ref{fig4}).  }
\label{fig1}
\end{figure*}

The \sca\ nuclear transition to its isomeric state at an energy of
$E_{\ind{0}}=12.389$~keV \cite{SRK23}, with a lifetime of
$\tau_{\ind{0}}=0.47$~s \cite{BHL67}, yields exceptionally small
natural width $\Gamma_{\ind{0}}=1.4$~feV and high quality factor
$Q_{\ind{0}}\simeq 10^{19}$—four orders of magnitude above the
operational quality factors of state-of-the-art optical clocks
\cite{Ludlow15}. This makes \sca\ a strong candidate for advanced
metrology and nuclear clocks. Resonant excitation of the \sca\ isomer
using x-ray free-electron laser (XFEL) pulses was recently
demonstrated \cite{SRK23}. Building on this success, we now
investigate the \sca\ resonance further and address a key question:
how closely do the actual linewidth $\Gamma$ and quality factor $Q$ of
the solid-state \sca\ resonance approach their natural limits?

\begin{table}[t!]
\centering
\begin{tabular}{ |l||c|c|c|c|c| } 
 \hline
Isomer  & $^{57}$Fe & $^{67}$Zn  & \thr & \sca & $^{109}$Ag \\ 
\hline
$E_{\ind{0}}$ (keV)  & 14.4 & 93.3  & 8.4$\cdot 10^{-3}$  & 12.4 & 88.0  \\
$\tau_{\ind{0}}$ (s)  & 1.4$\cdot 10^{-7}$ & 1.3$\cdot 10^{-5}$  & 641 \cite{ZOH24}  & 0.47 & 57.1  \\
$\Gamma_{\ind{0}}$ (eV)  & 4.8$\cdot 10^{-9}$ & 5$\cdot 10^{-11}$ & 1.0$\cdot 10^{-18}$ & 1.4$\cdot 10^{-15}$ & 1.2$\cdot 10^{-17}$    \\
$\Gamma_{\ind{0}}$ (Hz)  & 1.1$\cdot 10^{6}$ & 1.2$\cdot 10^{4}$    & 2.5$\cdot 10^{-4}$  & 0.34 & 2.9$\cdot 10^{-3}$ \\
$Q_{\ind{0}}$  & 3.1$\cdot 10^{12}$ & 1.9$\cdot 10^{15}$  & 8.1$\cdot 10^{18}$  & 8.8$\cdot 10^{18}$ & 7.5$\cdot 10^{21}$   \\
\hline
$\Gamma$ ($\Gamma_{\ind{0}}$)  & 1 & 1 \cite{PSS92}  & $10^8$ \cite{ODZ25}   & $>$ 5$\cdot 10^2$ \footnote{Present publication.}   & $\simeq 10$ \cite{BDI09}  \\
$Q=E_{\ind{0}}/\Gamma$  & 3$\cdot 10^{12}$ & 2$\cdot 10^{15}$  & $10^{11}$  & $< 10^{16}$  & $\simeq 10^{21}$
\\
 \hline
\end{tabular}
\caption{Selected nuclear isomers and
  their properties. The isomers are ordered by increasing value of the natural resonance quality factor $Q_{\ind{0}}$.}
\label{tab1}
\end{table}


{\em Approach--} Measuring resonance widths on the femto-eV scale is a
formidable challenge. A more practical approach is to observe the time
dependence of the nuclear resonant response on the complementary
millisecond scale.  To determine the actual linewidth $\Gamma$ of the
solid-state resonance, we aimed to detect coherent NFS from a \sca\
target, measure its time dependence, extract the decay time $\tau$,
and calculate $\Gamma=\hbar/\tau$ \cite{SS90}.  In contrast,
incoherent nuclear fluorescence reflects the natural lifetime
$\tau_0$, so confirming the small $\Gamma_0=\hbar/\tau_{\ind{0}}$ via
this method was also a goal. A further objective was to detect elastic
incoherent fluorescence at 12.4~keV—undetected in our earlier
experiment \cite{SRK23}—since it is essential for the future resonance
applications and its ratio to $K_{\alpha,\beta}$ emission yields the
partial $K$-shell internal conversion coefficient.

\begin{figure}[t!]
\includegraphics[width=0.49\textwidth]{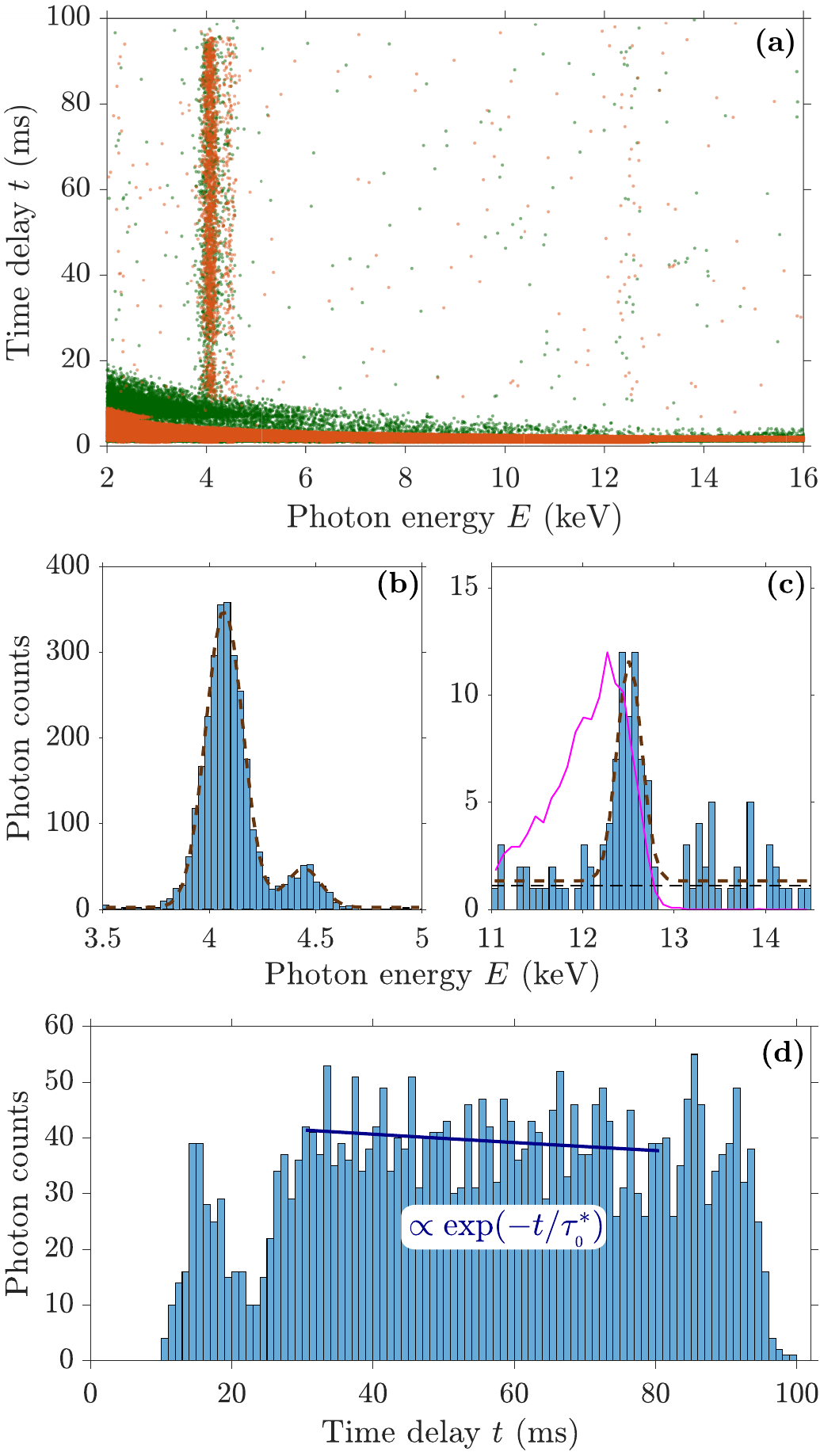}  
\caption{Incoherent fluorescence of \sca\ measured with x-ray
  detectors D$_{\indrm{d}}$ (red) and D$_{\indrm{u}}$ (green)
  following resonant excitation. (a) Detected events plotted as a
  function of re-emitted photon energy $E$ and time delay $t$ after
  excitation at $t=0$. (b)-(c) Photon counts from (a), integrated over
  15–100~ms plotted versus $E$ near 4~keV (b) and 12.4~keV (c); dashed
  lines indicate envelopes, and the solid magenta line in (c) shows
  the prompt Compton scattering profile. (d) $K_{\ind{\alpha,\beta}}$
  photon counts integrated over 3.75–4.75~keV, plotted as a function
  of $t$; the notch at 22~ms is an artifact due to shutter blade
  recoil.  }
  \label{fig2}
\end{figure}

{\em Experimental setup--} As in the first successful experiment
\cite{SRK23}, the \sca\ resonance was driven with 12.4-keV x-ray
pulses from the European XFEL (EuXFEL) in self-seeding mode
\cite{LGG23}, but now under improved conditions of three times higher
spectral flux. This enhancement enabled a more detailed investigation
of the resonance, including the observation of resonant elastic
fluorescence.

\begin{figure}[t!]
\includegraphics[width=0.5\textwidth]{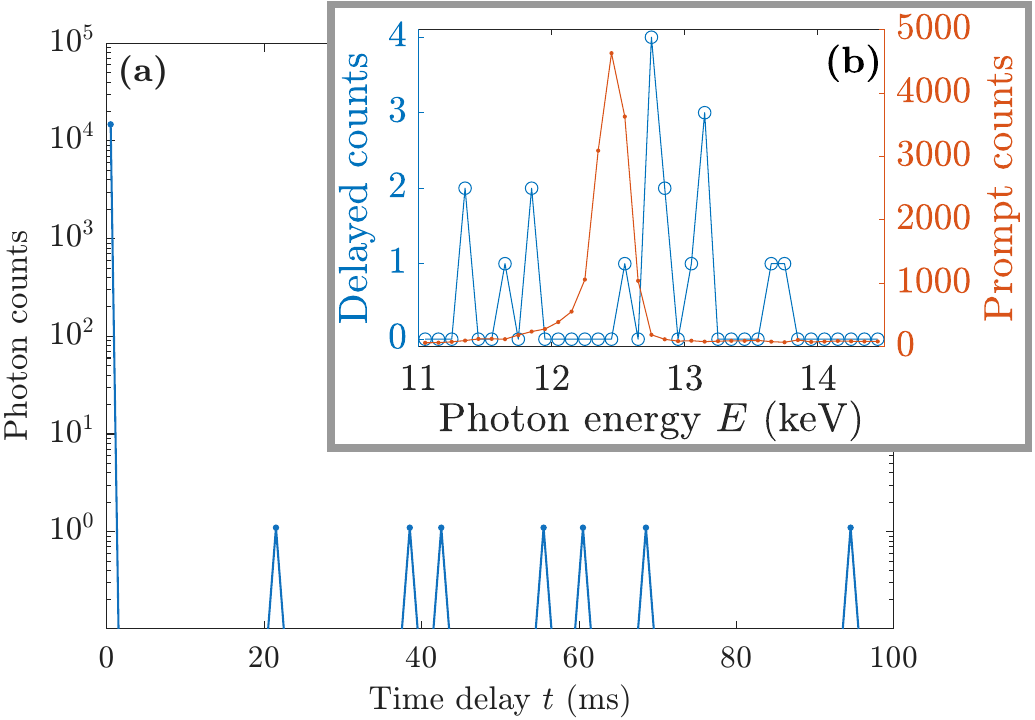}  
\caption{Photon counts recorded by  detector D$_{\mathrm{NFS}}$ in forward scattering geometry with  x-rays tuned to the \sca\ resonance in crystal targets Sc, ScN, \sco, or \sam. (a) Counts vs. time delay $t$, integrated over the {11.9–12.9}~keV photon energy range. (b) Counts vs. re-emitted energy $E$, separated into delayed (2–100~ms) and prompt (0–1~ms) windows.}
  \label{fig4}
\end{figure}

Figure~\ref{fig1} shows the experimental setup for resonant excitation
of \sca\ nuclei with 12.4-keV x-ray pulses and detection of both
coherent and incoherent time-delayed emission. Self-seeded XFEL
macropulses (pulse trains of 0.18~ms duration, much shorter than
$\tau_0$) excite \sca\ nuclei every 100~ms in two targets: first, a Sc
foil that shuttles within the \sca\ resonance-detection unit between
the x-ray beam and the detectors D${\indrm{u}}$ and D${\indrm{d}}$
\cite{SRK23}; second, a crystal target in the cryostat of the NFS
unit. Decay photons—elastic 12.4~keV and inelastic Sc
$K_{\alpha,\beta}$ fluorescence—are detected by Si drift detectors
D$_{\indrm{u}}$, D$_{\indrm{d}}$, and D$_{\indrm{NFS}}$ (Amptek
X123). Emission is measured as a function of energy $E$ (resolution
$<300$~eV) and time delay $t$ after pulse arrival ($t = 0$) with
$\mu$s resolution, as shown in Figs.~\ref{fig2} and \ref{fig4}. The
detectors' low background rate ($R_{\indrm{B}} =
0.9$~counts/keV/10,000~s) is essential for detecting the weak
\sca\ isomer decay signals.

The NFS unit is a new addition to the setup established in
\cite{SRK23}. It contains a set of NFS crystal targets (Sc, ScN, \sco,
and \sam) mounted on a cryofinger of a He-cryostat maintained at 20~K.
Cooling was important to avoid the resonance broadening due to
second-order Doppler effect, see discussion in \cite{SRK23}. Magnetic
field shielding, in the shape of a $\mu$-metal four-way cross centered
at the target, reduces the 50-$\mu$T earth field to 30~nT, effectively
eliminating externally induced magnetic splitting of the \sca\
resonance and NFS signal reduction.  The highest-quality single
crystals available were selected in order to minimize defect-induced
inhomogeneous broadening.  Only one NFS target is in the beam at a
time, raster-scanned to mitigate x-ray radiation load, and replaced
with another every $\simeq$1-2 hours. Two synchronous shutters with
0.25-mm thick W beamstops guard the detector D$_{\indrm{NFS}}$ from
direct x-ray excitation pulses and open the path for NFS photons to
D$_{\indrm{NFS}}$ 2~ms after excitation. This determines the smallest
measurable time delay in photon detection in the NFS experiment.

A Bond spectrometer is used to determine the absolute energy of the
incident photons and tune it to the known resonance energy of
12.38959~keV for \sca\ \cite{SRK23}. A minimally invasive single-shot
spectrometer \cite{KFL20} monitors x-ray pulse energy, relative photon
energy, and spectral bandwidth for each pulse.

\noindent {\em Results and discussion--} Figure~\ref{fig2}(a) shows
all photon events recorded with x-ray detectors in the resonance
detection unit D$_{\indrm{u}}$ (green dots) and D$_{\indrm{d}}$ (red
dots) as a function of re-emitted photon energy $E$ and time delay
$t$, while the incident x-rays were tuned to the \sca\ resonance. The
total number of macropulses was $\simeq 9\times 10^5$, and the data
acquisition time was $\simeq$25~hours.

Figures~\ref{fig2}(b)-(c) show photon counts integrated over the
15-100~ms time delay interval, plotted vs.\!\! energy~$E$ near 4~keV
(b) and 12.4~keV (c). Distinct $K_{\ind{\alpha}}$ and
$K_{\ind{\beta}}$ fluorescence lines appear in Fig.~\ref{fig2}(b). The
delayed $K_{\ind{\alpha,\beta}}$ photons count rate in both detectors
is $R_{\ind{4}}=328(6)$~ph/keV/10,000~s, yielding a signal-to-noise
ratio (SNR) of 183 -- nearly triple the previous SNR of 65
\cite{SRK23}, due to the narrower bandwidth (0.6 vs 1.3 eV) and higher
spectral flux of the self-seeded XFEL (see Methods).

Due to this improvement and longer acquisition time, we detected not
only $K_{\ind{\alpha,\beta}}$ fluorescence but also the much weaker
12.4-keV elastic delayed signal -- suppressed by internal electron
conversion (see Fig.~\ref{fig2}(c)). Its count rate is
$R_{\ind{12}}=7.3(0.9)$~ph/keV/10,000~s, with SNR$\simeq$4. The
magenta line in Fig.~\ref{fig2}(c) shows the prompt Compton
$\simeq 90^{\circ}$-scattering profile from the Sc target.

From the ratio $R_{\ind{4}}/R_{\ind{12}}$, accounting for the known
$K$-shell fluorescence yield $\omega_{\ind{K}}=0.19{0(2)}$
\cite{Krause79,HTS94} and geometrical factors, we determined the
partial internal conversion coefficient for the 12.4-keV to
ground-state transition in \sca\ to be $\alpha_{\ind{K}}=390(60)$ (see
Methods for details). This agrees with theoretical prediction
$\alpha_{\ind{K}} = 363$ \cite{KBT08} and the earlier indirect
estimate in \cite{SRK23}.

Figure~\ref{fig2}(d) shows incoherent $K_{\ind{\alpha,\beta}}$
fluorescence counts integrated over 3.75-4.75~keV and plotted versus
time delay $t$.   The notch at 22~ms is an artifact
  caused by the recoil of the shuttling blade carrying the Sc foil. 
Analysis of these data (see Methods) yields an
  isomer lifetime of $\tau_{\ind{0}}^*=0.46_{-0.1}^{+0.2}$~s. The
  relatively large uncertainty results from the limited time window of
  the measurements, which covers only 15\% of the 460-ms lifetime. The
  present data are consistent with a single-exponential decay, as
  expected for an isolated nuclear excited state, although their
  statistical precision does not allow the exponential character of
  the decay to be established independently. We therefore use the
  single-exponential model, supported by the previous
  Coulomb-excitation measurement [17], to extract the lifetime. The
  resulting value is consistent with the previously measured lifetime
  and, under this model, corresponds to a natural linewidth of
  $\Gamma_{\ind{0}}=1.4$~feV and an exceptionally high natural quality
  factor of $Q_{\ind{0}}\simeq10^{19}$ for the 12.4-keV \sca\
  isomer. 

Coherent NFS was measured simultaneously with incoherent
nuclear-resonant fluorescence. Figure~\ref{fig4} shows the time-delay
(a) and energy (b) spectra from detector D$_{\indrm{NFS}}$,
accumulated over 9.9~h for all NFS targets. Residual x-ray leakage
despite shutter protection ($\simeq 1$~ph/macropulse) produces a
prompt peak at $t=0$ and a line at $E=12.4$~keV. A weak feature near
$\simeq 12.6$~keV in Fig.~\ref{fig4}(b) is attributed to
cosmic-ray–induced Pb ($L_{\ind{\beta}}$) fluorescence in the
shielding. Within the 1-keV window set by the detector resolution, we
record total $N=7$ delayed counts, consistent with background and
showing no statistically significant excess.

Using low-count Poisson statistics, the observed $N=7$
{ counts in the NFS detection window correspond to a
  one-sided} 95\% confidence upper limit { obtained
  by inverting the cumulative Poisson distribution. Specifically, the
  upper limit $N_{\ind{95}}$ on the mean total number of counts
  satisfies $P(N\leq 7\,|\,N_{\ind{95}})=0.05$, giving
  $N_{\ind{95}}=13.14$, which we conservatively round to
  $N_{\ind{95}}\simeq 14$.}  Subtracting the expected background
$N_{\mathrm{bg}}=3$ within the observation window yields an upper
limit for the mean signal of $S_{95}=11$, corresponding to a mean
signal rate of $3$~ph/10,000~s.

To translate this experimental upper limit on the delayed coherent
signal into a constraint on the resonance linewidth, a model for the
coherent NFS response is required.  Accessing $\Gamma$ requires
detecting {\em coherent} resonance scattering from solid-state targets
\cite{SS90}, specifically through the time-dependent NFS rate $R(t)$
following prompt x-ray excitation:
\begin{equation} \label{eq001}
  \begin{split}
  R(t) & =2\pi \frac{N_{\ind{\Gamma_{\ind{0}}}}}{\tau_{\ind{0}}} \xi^2  \exp\left[-\left(\Gamma\!+\!\xi\Gamma_{\ind{0}}\right) t/\hbar-L/L_{\indrm{e}}\right],\\
  \xi &=L/4L_{\indrm{r}}, \hspace{1cm} L_{\indrm{r}}=1/(\sigma_{\indrm{r}} n_{\indrm{r}} f_{\indrm{ML}}).
\end{split}  
\end{equation}
Here $\xi$ is the nuclear resonance optical thickness parameter, $L$
is the thickness of the target, $L_{\indrm{r}}$ is the
nuclear-resonant absorption length, $\sigma_{\indrm{r}}$ is the
resonant cross-section, $n_{\indrm{r}}$ is the number density of
resonant nuclei, $f_{\indrm{ML}}$ is the Lamb–M\"ossbauer factor (the
probability of recoil-free elastic nuclear-resonant absorption and
emission, also known as the M\"ossbauer effect),
$N_{\ind{\Gamma_{\ind{0}}}}$ is the number of photons within the
natural linewidth $\Gamma_{\ind{0}}$ per incident x-ray pulse, and
$L_{\indrm{e}}$ is the photoelectric absorption
length. Equation~\eqref{eq001} is valid for $t \ll \tau_{\ind{0}}/\xi$
and for {a single resonance line with Lorentzian broadening}.

\begin{figure}[t!]
\includegraphics[width=0.5\textwidth]{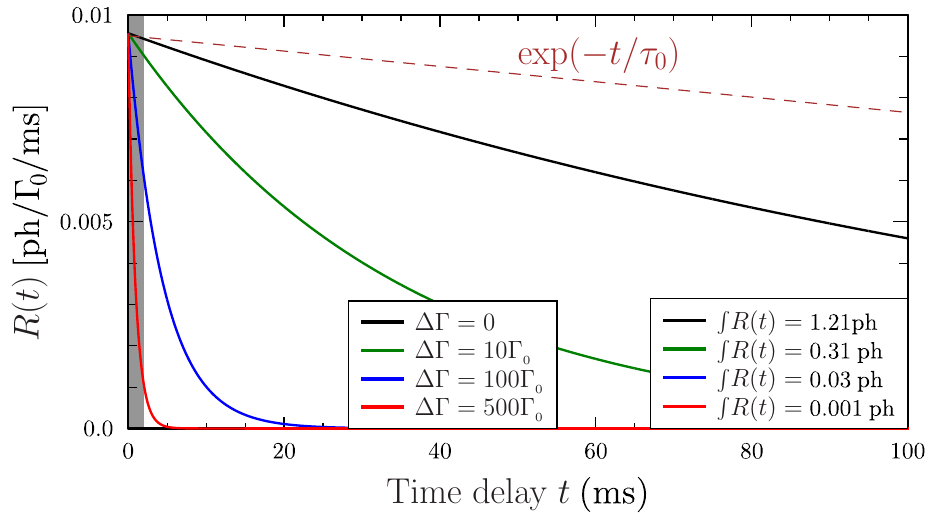}  
\caption{Nuclear forward scattering rate $R(t)$ of x-ray photons
  emitted by \sca\ nuclei as a function of the time delay $t$ after
  resonant excitation. Simulations were performed with
  the MOTIF code \cite{Sh00}, which yields results comparable to
  Eq.~\eqref{eq001} without approximations. Calculations assume a
  single resonance, optical thickness $\xi=2.25$ (as in the Sc, ScN,
  or \sco\ targets), inhomogeneous broadening $\Delta\Gamma$ of 0, 10,
  100, or 500~$\Gamma_{\ind{0}}$, and spectral density
  $N_{\ind{\Gamma_0}}=1$~ph/$\Gamma_{\ind{0}}$ per pulse. The dashed
  line indicates decay with the natural lifetime $\tau_{\indrm{0}}$;
  coherent NFS effects accelerate the decay even for $\Delta\Gamma=0$,
  while increasing $\Delta\Gamma$ further shortens the response and
  reduces its time-integrated strength. The inset shows $R(t)$
  integrated over the 2--100~ms time window for each $\Delta\Gamma$
  case. The time window $t<2$~ms, inaccessible in the experiment, is
  shown in gray. For the periodic excitation sequence used
  experimentally, $N_{\ind{\Gamma_0}}$ and the integrals correspond to
  photons per second rather than per pulse.}
  \label{fig3}
\end{figure}

The time dependence of $R(t)$ carries the fingerprint of inhomogeneous
broadening $\Gamma$, which results in a more rapid radiative decay in
the forward direction with a time constant $\tau=\hbar/\Gamma$ shorter
than the natural decay time $\tau_{\ind{0}}$ \cite{SS90}. The coherent
nature of NFS is evident in the quadratic enhancement
$R(t)\propto \xi^2$ and in the {\em homogeneous} resonance broadening
$\xi\Gamma_{\indrm{0}}$, which leads to an additional speedup of the
coherent nuclear decay \cite{KAK79,Kagan99,HT99,SSPT91,BSH92}.

For a spectral flux of
$\simeq0.3$~ph/$\Gamma_{\ind{0}}$/s on the NFS targets (see Methods),
simulations in Fig.~\ref{fig3} for
$\Delta\Gamma=500\,\Gamma_{\ind{0}}$ predict $\sim3$~ph/10,000~s,
approximately equal to the 95\% confidence upper limit on the mean NFS signal rate derived above.

We therefore infer that the coherent forward response decays before
the accessible delayed-time window, $t\lesssim 2$~ms. Within the
adopted single-line Lorentzian linewidth-broadening model and under
the present irradiation and excitation conditions, the null result
corresponds to an effective broadening
$\Delta\Gamma \gtrsim 500\,\Gamma_{\ind{0}}$ in the samples under
investigation, which includes all broadening and decoherence
mechanisms active during the measurement.  This provides a first
time-domain benchmark for decoherence and { effective} linewidth
broadening of the 12.4-keV \sca\ transition in condensed-matter hosts
at the femto-eV scale.

{ The numerical value of this lower bound is model-
and parameter-dependent, since it is obtained by comparing the observed
counts with simulations using the experimentally determined excitation
flux, target parameters, detection efficiency, background level, and
accessible delayed-time window.
However, the conclusion is qualitatively robust: within the adopted
model, coherent responses with linewidth broadening substantially
below this scale would have produced detectable delayed NFS
photons. Such long-lived responses are therefore excluded under the
conditions of this experiment and indicate broadening by more than two
orders of magnitude.}

The broadening may arise from several key mechanisms:\\ (1) Even in
perfect diamagnetic crystals, magnetic dipole–dipole interactions
between nuclear ground- and excited-state moments
($\vc{\mu}_{\indrm{g}}$, $\vc{\mu}_{\indrm{e}}$) cause inhomogeneous
energy shifts up to
$U \simeq 2{\mu}_{\indrm{g}}
{\mu}_{\indrm{e}}/r^3$, 
where $r$ is the inter-nuclear distance. This yields
$U \simeq 10^3 \Gamma_0$ for \sca\ in Sc and \sco\ or
$\simeq3~10^3\Gamma_0$ in ScN.  Note that for the \ag\ resonance $U$
was estimated to be $\approx 10^{5}\Gamma_{\ind{0}}^{\mathrm {Ag}}$
\cite{Davydov01}.  However, unexpectedly, the broadening claimed in
the experiment was only $\simeq 10\Gamma_{\ind{0}}^{\mathrm {Ag}}$
\cite{BDI09}.\\ (2) Nonzero electric field gradients (EFGs) in
non-cubic crystals cause quadrupole splitting of the nuclear states,
varying from $\sim 10^7 \Gamma_{\ind{0}}$ in Sc to
$\sim10^8 \Gamma_{\ind{0}}$ in \sco\ (see Methods). Defects modify
EFGs locally, inducing broadening. Of the tested crystals, only ScN
(cubic NaCl structure) should be free of this, but imperfect
stoichiometry likely introduced EFG-related effects.\\ (3)
Absorption of approximately 50~mJ per macropulse in
  the direct XFEL beam by the NFS targets may have caused transient
  electronic excitation and lattice distortions, leading to additional
  linewidth broadening. However, permanent radiation damage can be
excluded. First, the \sco\ crystals exhibited none of the
characteristic discoloration associated with radiation damage. Second,
x-ray Bragg diffraction topography performed on \sam\ crystals before
and after the experiment revealed no evidence of radiation-induced
damage within the sensitivity of the topographic technique.

{\em Summary and outlook --}  Resonant excitation and scattering from
\sca\ nuclei were observed by irradiating Sc targets with 12.4-keV
x-ray pulses at the European XFEL. 
{ Under very-low-background conditions, we measured
  incoherent delayed fluorescence and searched for coherent nuclear
  forward scattering (NFS).}

The \sca\ isomer lifetime was determined to be
$\taux=0.46_{-0.1}^{+0.2}$~s via Sc $K_{\alpha,\beta}$ fluorescence.
This confirms the persistence of the long-lived
  feV-narrow isomeric state in a solid-state environment following
  direct x-ray excitation — a scenario that had not been
  experimentally verified before.  It is a starting point for
  assessing the viability of \sca\ as a reference frequency standard
  with the extremely large $Q$ value in condensed matter systems.

  Elastic incoherent delayed fluorescence at 12.4 keV was observed
  confirming radiative decay as viable channel, essential for future
  narrow-band spectroscopy and nuclear clock development.  From the
  ratio of $K_{\alpha,\beta}$ to the elastic 12.4-keV line we obtained
  the partial $K$-shell internal conversion coefficient
  $\alpha_{\ind{K}}=390(60)$
  
We introduce a femto-eV-resolution, time-domain
  probe of the solid-state \sca\ linewidth based on coherent nuclear
  forward scattering and { use it to place an experimental lower bound
  on the broadening of the \sca\ isomeric resonance. Under the present
  experimental conditions, the absence of a detectable delayed
  coherent forward response beyond 2~ms leads us, within the adopted
  linewidth-broadening model, to infer} an effective broadening
  $\Delta\Gamma \gtrsim 500\,\Gamma_{\ind{0}}$.
  This experiment
  provides the first quantitative constraint on { the
    linewidth broadening of a solid-state nuclear transition with an
    intrinsic natural linewidth in the femto-electronvolt range}.  It
  opens a new frontier with implications for x-ray nuclear clocks,
  precision metrology, and quantum technologies.

Advancing this research will require reducing resonance broadening by
improving NFS-target crystal quality, 
    increasing spectral flux, minimizing x-ray–induced damage,
accessing shorter delay times, and applying dynamic nuclear
resonance–narrowing techniques \cite{IK74,AIK75,ARK07}.

Successful detection of \sca\ NFS would enable ultra-precise
spectroscopy at feV resolution in the hard x-ray regime and guide the
choice of host materials and excitation/readout protocols for
nuclear-clock applications.

Both the \sca\ and \thr\ nuclear frequency standards present viable
yet complementary challenges. The \thr\ isomeric transition, lying in
the optical range (8.4~eV), benefits from existing high-precision
laser and frequency-comb technologies \cite{ZOH24}, but achieving its
ultimate quality factor ($Q \simeq 10^{19}$) is hindered by severe
solid-state broadening and environmental sensitivity, with current
experiments limited to $Q \simeq 10^{11}$\cite{ODZ25}. In contrast,
the \sca\ transition at 12.4 keV { has three orders of magnitude higher frequency,} is intrinsically less sensitive to
its environment, and thus more likely to preserve a high $Q$ even in
solids (projected $Q \simeq 10^{16}$ before applying
resonance-narrowing techniques), but its realization depends on the
development of coherent X-ray frequency-comb sources, such as
intracavity high-harmonic generation \cite{HHG1,HHG2,FCXUV1Hz,GWL22}
or X-ray free-electron laser oscillators
\cite{KSR08,AK15,RBF26}. Consequently, \thr\ is constrained mainly by
materials physics, while \sca\ is limited by photonic technology,
making both approaches challenging yet promising and ultimately
complementary paths toward nuclear-based time and frequency standards.

{\em Acknowledgments--} 
This research used resources of the Advanced Photon
Source, a U.S. Department of Energy (DOE) Office of Science user
facility at Argonne National Laboratory (ANL) and is based on research
supported by the U.S. DOE Office of Science--Basic Energy Sciences,
under Contract No. DE-AC02-06CH11357.
Work at Texas A\&M University and ANL was also supported by the
National Science Foundation (grant No. PHY-2409734``New Horizons in
Quantum Nuclear X-ray Optics'').  The synthesis and crystal growth of
ScN at Kansas State University was supported by the National Science
Foundation (grant No. 1508172).  This work was supported and partly
funded by the Cluster of Excellence “Advanced Imaging of Matter” of
the Deutsche Forschungsgemeinschaft (DFG)—EXC 2056—project ID
390715994. T.K. acknowledges support from Ministry of Science and Higher Education, Poland, under contract no. 1/SOL/2021/2.

We acknowledge European XFEL in Schenefeld, Germany, for provision of
X-ray free-electron laser beamtime at the Materials Imaging and
Dynamics (MID) instrument located at the SASE-2 beamline and would
like to thank the staff for their assistance, in particular Uwe Englisch, James Moore,
Andrea Parenti, and  James Wrigley. Data recorded for 
experiment \#6536 at the European XFEL  are available at the
https://in.xfel.eu/metadata/doi/10.22003/XFEL.EU-DATA-006536-00.

Competing interests: The authors declare no competing interests.


\begin{thebibliography}{10}

\bibitem{Ludlow15}
Andrew~D. Ludlow, Martin~M. Boyd, Jun Ye, E.~Peik, and P.~O. Schmidt.
\newblock Optical atomic clocks.
\newblock {\em Rev. Mod. Phys.}, 87:637--701, Jun 2015.

\bibitem{SBD18}
M.~S. Safronova, D.~Budker, D.~DeMille, Derek F.~Jackson Kimball,
  A.~Derevianko, and Charles~W. Clark.
\newblock Search for new physics with atoms and molecules.
\newblock {\em Rev. Mod. Phys.}, 90:025008, Jun 2018.

\bibitem{BSS21}
Kjeld Beeks, Tomas Sikorsky, Thorsten Schumm, Johannes Thielking, Maxim~V.
  Okhapkin, and Ekkehard Peik.
\newblock The thorium-229 low-energy isomer and the nuclear clock.
\newblock {\em Nat. Rev. Phys.}, 3:238--248, 2021.

\bibitem{PSS21}
E~Peik, T~Schumm, M~S Safronova, A~Pálffy, J~Weitenberg, and P~G Thirolf.
\newblock Nuclear clocks for testing fundamental physics.
\newblock {\em Quantum Science and Technology}, 6(3):034002, apr 2021.

\bibitem{Wertheim64}
Gunther~K. Wertheim.
\newblock {\em {M}\"ossbauer {E}ffect {P}rinciples and {A}pplications}.
\newblock Academic Press New York and London, 1964.
\newblock ISBN 978-1-4832-2856-3.

\bibitem{PSS92}
W.~Potzel, C.~Sch{\"a}fer, M.~Steiner, H.~Karzel, W.~Schiessl, M.~Peter, G.~M.
  Kalvius, T.~Katila, E.~Ikonen, P.~Helist{\"o}, J.~Hietaniemi, and K.~Riski.
\newblock Gravitational redshift experiments with the high-resolution
  {M}{\"o}ssbauer resonance in $^{67}${Zn}.
\newblock {\em Hyperfine Interact.}, 72:195--214, 1992.

\bibitem{BDI09}
Yu.~D. Bayukov, A.~V. Davydov, Yu.~N. Isaev, G.~R. Kartashov, M.~M. Korotkov,
  and V.~V. Migachev.
\newblock Observation of the gamma resonance of a long-lived $^{109m}${Ag}
  isomer using a gravitational gamma-ray spectrometer.
\newblock {\em JETP Lett.}, 90:499--503, 2009.

\bibitem{WSL16}
Lars von~der Wense, Benedict Seiferle, Mustapha Laatiaou, J\"urgen~B. Neumayr,
  Hans-J\"org Maier, Hans-Friedrich Wirth, Christoph Mokry, J\"org Runke, Klaus
  Eberhard, Christoph~E. D\"ullmann, Norbert~G. Trautmann, and Peter~G.
  Thirolf.
\newblock Direct detection of the $^{229}${Th} nuclear clock transition.
\newblock {\em Nature}, 533:47--51, 2016.

\bibitem{Thfreq20}
Tomas Sikorsky, Jeschua Geist, Daniel Hengstler, Sebastian Kempf, Loredana
  Gastaldo, Christian Enss, Christoph Mokry, J\"org Runke, Christoph~E.
  D\"ullmann, Peter Wobrauschek, Kjeld Beeks, Veronika Rosecker, Johannes~H.
  Sterba, Georgy Kazakov, Thorsten Schumm, and Andreas Fleischmann.
\newblock Measurement of the $^{229}\mathrm{Th}$ isomer energy with a magnetic
  microcalorimeter.
\newblock {\em Phys. Rev. Lett.}, 125:142503, 2020.

\bibitem{SWB19}
Benedict Seiferle, Lars von~der Wense, Pavlo~V. Bilous, Ines Amersdorffer,
  Christoph Lemell, Florian Libisch, Simon Stellmer, Thorsten Schumm,
  Christoph~E. Düllmann, Adriana P{\'a}lffy, and Peter~G. Thirolf.
\newblock Energy of the $^{229}${Th} nuclear clock transition.
\newblock {\em Nature}, 573:243--246, 2019.

\bibitem{KMA23}
Sandro Kraemer, Janni Moens, Michail Athanasakis-Kaklamanakis, Silvia Bara,
  Kjeld Beeks, Premaditya Chhetri, Katerina Chrysalidis, Arno Claessens,
  Thomas~E. Cocolios, João G.~M. Correia, Hilde~De Witte, Rafael Ferrer,
  Sarina Geldhof, Reinhard Heinke, Niyusha Hosseini, Mark Huyse, Ulli
  K{\"o}ster, Yuri Kudryavtsev, Mustapha Laatiaoui, Razvan Lica, Goele
  Magchiels, Vladimir Manea, Clement Merckling, Lino M.~C. Pereira, Sebastian
  Raeder, Thorsten Schumm, Simon Sels, Peter~G. Thirolf, Shandirai~Malven
  Tunhuma, Paul Van Den~Bergh, Piet Van~Duppen, Andr\'e Vantomme, Matthias
  Verlinde, Renan Villarreal, and Ulrich Wahl.
\newblock {Observation of the radiative decay of the $^{229}${Th} nuclear clock
  isomer}.
\newblock {\em Nature}, 617:706--710, May 2023.

\bibitem{TOZ24}
J.~Tiedau, M.~V. Okhapkin, K.~Zhang, J.~Thielking, G.~Zitzer, E.~Peik,
  F.~Schaden, T.~Pronebner, I.~Morawetz, L.~Toscani De~Col, F.~Schneider,
  A.~Leitner, M.~Pressler, G.~A. Kazakov, K.~Beeks, T.~Sikorsky, and T.~Schumm.
\newblock Laser excitation of the {Th}-229 nucleus.
\newblock {\em Phys. Rev. Lett.}, 132:182501, Apr 2024.

\bibitem{ESJ24}
R.~Elwell, Christian Schneider, Justin Jeet, J.~E.~S. Terhune, H.~W.~T. Morgan,
  A.~N. Alexandrova, H.~B. Tran~Tan, Andrei Derevianko, and Eric~R. Hudson.
\newblock Laser excitation of the $^{229}\mathrm{Th}$ nuclear isomeric
  transition in a solid-state host.
\newblock {\em Phys. Rev. Lett.}, 133:013201, Jul 2024.

\bibitem{ZOH24}
Chuankun Zhang, Tian Ooi, Jacob~S. Higgins, Jack~F. Doyle, Lars von~der Wense,
  Kjeld Beeks, Adrian Leitner, Georgy~A. Kazakov, Peng Li, Peter~G. Thirolf,
  Thorsten Schumm, and Jun Ye.
\newblock Frequency ratio of the $^{229m}${Th} nuclear isomeric transition and
  the $^{87}${Sr} atomic clock.
\newblock {\em Nature}, 633:63--70, 2024.

\bibitem{ODZ25}
Tian Ooi, Jack~F. Doyle, Chuankun Zhang, Jacob~S. Higgins, Jun Ye, Kjeld Beeks,
  Tomas Sikorsky, and Thorsten Schumm.
\newblock Frequency reproducibility of solid-state {Th-229} nuclear clocks.
\newblock July 2025.
\newblock arXiv:2507.01180v1.

\bibitem{SRK23}
Yuri Shvyd'ko, Ralf R\"ohlsberger, Olga Kocharovskaya, J\"org Evers,
  Gianluca~Aldo Geloni, Peifan Liu, Deming Shu, Antonino Miceli, Brandon Stone,
  Willi Hippler, Berit Marx-Glowna, Ingo Uschmann, Robert Loetzsch, Olaf
  Leupold, Hans-Christian Wille, Ilya Sergeev, Miriam Gerharz, Xiwen Zhang,
  Christian Grech, Marc Guetg, Vitali Kocharyan, Naresh Kujala, Shan Liu,
  Weilun Qin, Alexey Zozulya, J\"org Hallmann, Ulrike Boesenberg, Wonhyuk Jo,
  Johannes M\"oller, Angel Rodriguez-Fernandez, Mohamed Youssef, Anders Madsen,
  and Tomasz Kolodziej.
\newblock Resonant x-ray excitation of the nuclear clock isomer $^{45}${Sc}.
\newblock {\em Nature}, 622:471--475, 2023.

\bibitem{BHL67}
A.~E. Blaugrund, R.~E. Holland, and F.~J. Lynch.
\newblock Coulomb excitation of low-lying excited states in {Sc}$^{45}$.
\newblock {\em Phys. Rev.}, 159:926--930, Jul 1967.

\bibitem{SS90}
{Yu}.~V. Shvyd'ko and G.~V. Smirnov.
\newblock On the direct measurement of nuclear $\gamma$-resonance parameters of
  long-lived ($\gtrsim 1$~s) isomers.
\newblock {\em Nucl. Instrum. Methods Phys. Res. B}, 51:452--457, 1990.

\bibitem{LGG23}
Shan Liu, Christian Grech, Marc Guetg, Suren Karabekyan, Vitali Kocharyan,
  Naresh Kujala, Christoph Lechner, Tianyun Long, Najmeh Mirian, Weilun Qin,
  Svitozar Serkez, Sergey Tomin, Jiawei Yan, Suren Abeghyan, Jayson Anton,
  Vladimir Blank, Ukrike Boesenberg, Frank Brinker, Ye~Chen, Winfried Decking,
  Xiaohao Dong, Steve Kearney, Daniele~La Civita, Anders Madsen, Theophilos
  Maltezopoulos, Angel Rodriguez-Fernandez, Evgueni Saldin, Liubov Samoylova,
  Matthias Scholz, Harald Sinn, Vivien Sleziona, Deming Shu, Takanori Tanikawa,
  Sergey Terentyev, Andrei Trebushinin, Thomas Tschentscher, Maurizio Vannoni,
  Torsten Wohlenberg, Mikhail Yakopov, and Gianluca Geloni.
\newblock Cascaded hard x-ray self-seeded free-electron laser at
  {MHz}-repetition-rate.
\newblock {\em Nature Photonics}, 2023.

\bibitem{KFL20}
Naresh Kujala, Wolfgang Freund, Jia Liu, Andreas Koch, Torben Falk, Marc
  Planas, Florian Dietrich, Joakim Laksman, Theophilos Maltezopoulos, Johannes
  Risch, Fabio Dall'Antonia, and Jan Gr{\"u}nert.
\newblock Hard x-ray single-shot spectrometer at the {European} {X-ray}
  {Free}-{Electron} {Laser}.
\newblock {\em Rev. Sci. Instrum.}, 91(10):103101, 2020.

\bibitem{Krause79}
M.~O. Krause.
\newblock Atomic radiative and radiationless yields for {$K$} and {$L$} shells.
\newblock {\em Journal of Physical and Chemical Reference Data}, 8(2):307--327,
  1979.

\bibitem{HTS94}
J.~H. Hubbell, P.~N. Trehan, Nirmal Singh, B.~Chand, D.~Mehta, M.~L. Garg,
  R.~R. Garg, Surinder Singh, and S.~Puri.
\newblock A review, bibliography, and tabulation of {$K$}, {$L$}, and higher
  atomic shell {X-}ray fluorescence yields.
\newblock {\em Journal of Physical and Chemical Reference Data},
  23(2):339--364, 1994.

\bibitem{KBT08}
T.~Kib{\'e}di, T.W. Burrows, M.B. Trzhaskovskaya, P.M. Davidson, and C.W.
  Nestor.
\newblock Evaluation of theoretical conversion coefficients using {BrIcc}.
\newblock {\em Nucl. Instrum. Methods Phys. Res. A}, 589(2):202--229, 2008.

\bibitem{Sh00}
{Yu}.~V. Shvyd'ko.
\newblock {MOTIF}: {Evaluation} of time spectra for nuclear forward scattering.
\newblock {\em Hyperfine Interact.}, 125:173--188, 2000.

\bibitem{KAK79}
{Yu}. Kagan, A.~M. Afanas'ev, and V.~G. Kohn.
\newblock On excitation of isomeric nuclear states in a crystal by synchrotron
  radiation.
\newblock {\em J. Phys. C: Solid St. Phys.}, 12:615--631, 1979.

\bibitem{Kagan99}
Yu. Kagan.
\newblock Theory of coherent phenomena and fundamentals in nuclear resonant
  scattering.
\newblock {\em Hyperfine Interactions}, 123:83--126, 1999.

\bibitem{HT99}
J.~P. Hannon and G.~T. Trammell.
\newblock Coherent $\gamma$-ray optics.
\newblock {\em Hyperfine Interact.}, 123/124:127--274, 1999.

\bibitem{SSPT91}
{Yu}.~V. Shvyd'ko, G.~V. Smirnov, S.~L. Popov, and T.~Hertrich.
\newblock Observation of the intensified forward $\gamma$-ray emission in
  spontaneous nuclear decay.
\newblock {\em Pis'ma. Zh. Eksp. Teor. Fiz.}, 53:69, Jan 1991.
\newblock [(1991) {\em JETP Lett.} {\bf 53}, 69 ].

\bibitem{BSH92}
U.~van B{\"u}rck, D.~P. Siddons, J.~B. Hastings, U.~Bergmann, and R.~Hollatz.
\newblock Nuclear forward scattering of synchrotron radiation.
\newblock {\em Phys. Rev. B}, 46:6207, 1992.

\bibitem{Davydov01}
Andrey~V. Davydov.
\newblock The gamma resonance problem of long-lived nuclear isomers.
\newblock {\em Hyperfine Interactions}, 135:125--153, 2001.

\bibitem{IK74}
Yu.~A. Il'inskii and R.V. Khokhlov.
\newblock Narrowing of gamma resonance lines in crystals by radio-frequency
  fields.
\newblock {\em JETP}, 38(4):809--812, 1974.
\newblock Russian original - ZhETF, Vol. 65, No. 4, p. 1619, April 1974.

\bibitem{AIK75}
A.V. Andreev, Yu.~A. Il'inskii, and R.V. Khokhlov.
\newblock Narrowing of gamma resonance lines in crystals by continuous
  radio-frequency fields.
\newblock {\em JETP}, 40(5):819--820, 1975.
\newblock Russian original - ZhETF, Vol. 67, No. 5, p. 1647, March 1975.

\bibitem{ARK07}
Petr Anisimov, Yuri Rostovtsev, and Olga Kocharovskaya.
\newblock Concept of spinning magnetic field at magic-angle condition for line
  narrowing in m\"ossbauer spectroscopy.
\newblock {\em Phys. Rev. B}, 76:094422, Sep 2007.

\bibitem{HHG1}
R.J. Jones, K.D. Moll, M.J. Thorpe, and Ye~Jun.
\newblock Phase-coherent frequency combs in the vacuum ultraviolet via
  high-harmonic generation inside a femtosecond enhancement cavity.
\newblock {\em Phys. Rev. Lett.}, 94:93201--932014, May 2005.

\bibitem{HHG2}
A.~Cing\"oz, Dylan~C. Yost, T.~K. Allison, A.~Ruehl, M.~E. Fermann, I.~Hartl,
  and Ye~Jun.
\newblock Direct frequency comb spectroscopy in the extreme ultraviolet.
\newblock {\em Nature}, 482:68--71, February 2012.

\bibitem{FCXUV1Hz}
Craig Benko, Thomas~K. Allison, Arman Cingo, Linqiang Hua, Francois Labaye,
  Dylan~C. Yost, and Jun Ye.
\newblock Extreme ultraviolet radiation with coherence time greater than 1s.
\newblock 8:530--537, 2014.

\bibitem{GWL22}
Jixing Gao, Jiaqi Wu, Zhiyuan Lou, Fan Yang, Junyu Qian, Yujie Peng, Yuxin
  Leng, Yinghui Zheng, Zhinan Zeng, and Ruxin Li.
\newblock High-order harmonic generation in an x-ray range from laser-induced
  multivalent ions of noble gas.
\newblock {\em Optica}, 9(9):1003--1008, Sep 2022.

\bibitem{KSR08}
Kwang-Je Kim, Yuri Shvyd\char39{}ko, and Sven Reiche.
\newblock A proposal for an x-ray free-electron laser oscillator with an
  energy-recovery linac.
\newblock {\em Phys. Rev. Lett.}, 100(24):244802, Jun 2008.

\bibitem{AK15}
B.~W. Adams and K.-J. Kim.
\newblock X-ray comb generation from nuclear-resonance-stabilized x-ray
  free-electron laser oscillator for fundamental physics and precision
  metrology.
\newblock {\em Phys. Rev. ST Accel. Beams}, 18:030711, Mar 2015.

\bibitem{RBF26}
Patrick Rauer, Immo Bahns, Bertram Friedrich, Sara Casalbuoni, Massimiliano~Di
  Felice, Martin Dommach, Idoia~Freijo Martin, Wolfgang Freund, Jan Grünert,
  Marc Guetg, Ivars Karpics, Suren Karabekyan, Andreas Koch, Naresh Kujala,
  Daniele~La Civita, Jia Liu, Theophilos Maltezopoulos, Mikako Makita, Frank
  Mayet, Lukas Müller, Benoit Rio, Liubov Samoylova, Silja Schmidtchen,
  Matthias Scholz, Alessandro Silenzi, Vivienne Strauch, Daniel Thoden, Torsten
  Wohlenberg, Maurizio Vannoni, Fan Yang, Winfried Decking, Joerg Rossbach, and
  Harald Sinn.
\newblock Lasing of a cavity-based {X}-ray source.
\newblock {\em Nature}, 650:93--96, 2026.

\bibitem{MHA21}
A.~Madsen, J.~Hallmann, G.~Ansaldi, T.~Roth, W.~Lu, C.~Kim, U.~Boesenberg,
  A.~Zozulya, J.~M{\"{o}}ller, R.~Shayduk, M.~Scholz, A.~Bartmann, A.~Schmidt,
  I.~Lobato, K.~Sukharnikov, M.~Reiser, K.~Kazarian, and I.~Petrov.
\newblock {Materials Imaging and Dynamics (MID) instrument at the European
  X-ray Free-Electron Laser Facility}.
\newblock {\em Journal of Synchrotron Radiation}, 28(2):637--649, Mar 2021.

\bibitem{GML10}
Benjamin~K Greve, Kenneth~L. Martin, Peter~L. Lee, Peter~J. Chupas, Karena~W.
  Chapman, and Angus~P. Wilkinson.
\newblock {Pronounced Negative Thermal Expansion from a Simple Structure: Cubic
  ScF$_3$}.
\newblock {\em J. Am. Chem. Soc.}, 132:15498, 2010.

\bibitem{KBA19}
Denis Karimov, Irina Buchinskaya, Natalia Arkharova, Pavel Prosekov, Vadim
  Grebenev, Nikolay Sorokin, Tatiana Glushkova, and Pavel Popov.
\newblock Growth from the melt and properties investigation of {ScF}$_3$ single
  crystals.
\newblock {\em Crystals}, 9(7):371, Jul 2019.

\bibitem{ISF21}
Katsuhiko Inaba, Kazumasa Sugiyama, Takashi Fujii, and Tsuguo Fukuda.
\newblock X-ray diffraction analysis and x-ray topography of high-quality
  {ScAlMgO}$_4$ substrates.
\newblock {\em Journal of Crystal Growth}, 574:126322, 2021.

\bibitem{GMZ06}
Y.~Gangrsky, K.~Marinova, S.~Zemlyanoi, M.~Avgoulea, J.~Billowes, P.~Campbell,
  B.~Cheal, B.~Tordoff, M.~Bissell, D.~H. Forest, M.~Gardner, G.~Tungate,
  J.~Huikari, H.~Penttil{\"a}, and J.~{\"A}yst{\"o}.
\newblock Nuclear charge radii and electromagnetic moments of scandium isotopes
  and isomers in the $f_{7/2}$ shell.
\newblock {\em Hyperfine Interact}, 171:209--215, 2006.

\bibitem{BBS65}
R.~G. Barnes, F.~Borsa, S.~L. Segel, and D.~R. Torgeson.
\newblock Knight shift anisotropy in scandium and yttrium and nuclear
  quadrupole coupling in scandium.
\newblock {\em Phys. Rev.}, 137:A1828--A1834, Mar 1965.

\bibitem{RFI69}
J.~W. Ross, F.~Y. Fradin, L.~L. Isaacs, and D.~J. Lam.
\newblock Magnetic and nuclear-resonance properties of single-crystal scandium.
\newblock {\em Phys. Rev.}, 183:645--652, Jul 1969.

\bibitem{CR57}
M.H. Cohen and F.~Reif.
\newblock Quadrupole effects in nuclear magnetic resonance studies of solids.
\newblock volume~5 of {\em Solid State Physics}, pages 321--438. Academic
  Press, 1957.

\bibitem{SEB22}
Jennifer Steinadler, Lucien Eisenburger, and Thomas Bräuniger.
\newblock Characterization of the binary nitrides {VN} and {ScN} by solid-state
  {NMR} spectroscopy.
\newblock {\em Zeitschrift f\"ur anorganische und allgemeine Chemie},
  648(21):e202200201, 2022.

\bibitem{KRZ07}
Dzhalil Khabibulin, Konstantin Romanenko, Mikhail Zuev, and Olga Lapina.
\newblock Solid state {NMR} characterization of individual compounds and solid
  solutions formed in {Sc$_2$O$_3$-V$_2$O$_5$-Nb$_2$O$_5$-Ta$_2$O$_5$} system.
\newblock {\em Magnetic Resonance in Chemistry}, 45(11):962--970, 2007.

\bibitem{OT14}
Itaru Oikawa and Hitoshi Takamura.
\newblock $^{45}${Sc} {NMR} spectroscopy and first-principles calculation on
  the symmetry of {ScO}$_6$ polyhedra in {BaO–Sc}$_2${O}$_3$-based oxides.
\newblock {\em Dalton Trans.}, 43:9714--9721, 2014.

\bibitem{AZC20}
Hayder Al-Atabi, Qiye Zheng, John~S. Cetnar, David Look, David~G. Cahill, and
  James~H. Edgar.
\newblock {Properties of bulk scandium nitride crystals grown by physical vapor
  transport}.
\newblock {\em Applied Physics Letters}, 116(13):132103, 04 2020.

\bibitem{MMS23}
Takashi Matsuoka, Hitoshi Morioka, Satoshi Semboshi, Yukihiko Okada, Kazuya
  Yamamura, Shigeyuki Kuboya, Hiroshi Okamoto, and Tsuguo Fukuda.
\newblock Properties of {ScAlMgO}$_4$ as substrate for nitride semiconductors.
\newblock {\em Crystals}, 13(3), 2023.

\bibitem{PBP02}
V.~Peters, A.~Bolz, K.~Petermann, and G.~Huber.
\newblock Growth of high-melting sesquioxides by the heat exchanger method.
\newblock {\em Journal of Crystal Growth}, 237-239:879--883, 2002.

\bibitem{AZH22}
Hayder~A. Al-Atabi, Xiaotian Zhang, Shanmei He, Cheng chen, Yulin Chen, Eli
  Rotenberg, and James~H. Edgar.
\newblock Lattice and electronic structure of {ScN} observed by angle-resolved
  photoemission spectroscopy measurements.
\newblock {\em Applied Physics Letters}, 121(18):182102, 11 2022.

\bibitem{STB21}
Yuri Shvyd'ko, Sergey Terentyev, Vladimir Blank, and Tomasz Kolodziej.
\newblock {Diamond channel-cut crystals for high-heat-load beam-multiplexing
  narrow-band X-ray monochromators}.
\newblock {\em Journal of Synchrotron Radiation}, 28(6):1720--1728, Nov 2021.

\bibitem{TPD22}
K~R Tasca, I~Petrov, C~Deiter, S~Martyushov, S~Polyakov, A~Rodriguez-Fernandez,
  R~Shayduk, H~Sinn, S~Terentyev, and M~Vannoni.
\newblock Study of a diamond channel cut monochromator for high repetition rate
  operation at the {EuXFEL}: {FEA} thermal load simulations and first
  experimental results.
\newblock {\em J. Phys.: Conf. Ser.}, 2380:012053, 2022.

\end{thebibliography}

\newpage
\appendix
\section{\large Methods}

\noindent
    {\bf Spectral flux.} The experiment was performed at the Materials
    Imaging and Dynamics (MID) instrument at the EuXFEL \cite{MHA21}.
    For this experiment, the European XFEL delivered x-ray pulses at
    the undulator exit with pulse energy $E_{\indrm{p}} = 0.55$~mJ, of
    which $E_{\indrm{bg}}$=0.08-mJ is SASE background, bandwidth
    $\Delta E_{\indrm{p}} = 0.6$~eV, and spectral density
    $S_{\indrm{p}} = (E_{\indrm{p}}-E_{\indrm{bg}})/\Delta
    E_{\indrm{p}} = 0.78$~mJ/eV = $5.5 \times
    10^{-4}$~ph/$\Gamma_{\ind{0}}$. This is about three times higher
    than in the first experiment \cite{SRK23}, where $S_{\indrm{p}}^*
    = 0.27$~mJ/eV with $E_{\indrm{p}}^* = 0.35$~mJ and $\Delta
    E_{\indrm{p}}^* = 1.3$~eV.

The pulses were delivered in trains of $n_{\indrm{p}} = 400$, spaced
by 440~ns, with a total train duration of 0.18~ms at a 10~Hz
repetition rate. Since this duration is much shorter than the
\sca\ isomer lifetime $\tau_{\ind{0}}$, the train is treated as a
single macropulse in the first approximation.

Under these conditions, the spectral flux at the undulator exit is $F
= 10 S_{\indrm{p}} n_{\indrm{p}} \simeq
2.2$~ph/$\Gamma_{\ind{0}}$/s. Accounting for a cumulative transmission
factor of 0.44 through beamline optics, the flux at the resonance
detection unit is $F_{\indrm{RDU}} \simeq 1$~ph/s/$\Gamma_{\ind{0}}$.

The estimated spectral flux at the NFS target is $F_{\indrm{NFS}} =$
$F_{\indrm{RDU}} T \simeq 0.3$~ph/s/$\Gamma_{\ind{0}}$, where the
total transmission $T = T_{\indrm{Sc}} T_{\indrm{air}} T_{\indrm{CVD}}
T_{\indrm{GC}} = 0.3$ includes contributions from subsequent beamline
components at the MID instrument: $T_{\indrm{Sc}} = 0.66$ (25-$\mu$m
Sc foil), $T_{\indrm{air}} = 0.7$ (0.75~m air path), $T_{\indrm{CVD}}
= 0.75$ (700-$\mu$m CVD diamond), and $T_{\indrm{GC}} = 0.87$
(800-$\mu$m glassy carbon window of the cryostat).

The NFS shutters and the NFS detectors were iteratively mounted into
the forward beam using an optical alignment laser, burn paper and
variable x-ray attenuators for alignment measurements with open
shutters.

\begin{table*}[t!]
\begin{tabular}{||l||l|l|l|l|l||} 
 \hline         \hline        
 crystal &   Sc &   ScN  &   \sco\  &   \scf\ &   \sam  \\
 &     & (Sc$_4$N$_4$) & (Sc$_8$Sc$_{24}$O$_{48}$) & (Sc$_1$F$_3$) & (Sc$_{3}$Al$_{3}$Mg$_{3}$O$_{12}$) \\
   &  &  &   & & \\
 \hline         \hline         
 space group    & $P63/mmc$  	 & $Fm\overline{3}m$   & $Ia\overline{3}$ 	 &   $Pm\overline{3}m$ & $R\overline{3}m$ \\   
                & (194) hcp	                 & (225) NaCl                    & (206)	 &  (221) \cite{GML10,KBA19} & (166) \cite{ISF21}\\
\hline         
magnetism  & paramagnetic & diamagnetic & diamagnetic & diamagnetic & diamagnetic \\
\hline
Quadrupole interaction  parameter &  &  &   &  & \\
$eQ_{\indrm{g}} V_{\ind{zz}}/h$ \footnote{The quadrupole interaction parameters for the excited nuclear states $eQ_{\indrm{e}} V_{\ind{zz}}/h$ are derived from the appropriate ground state values using the ratio of the  \sca\ quadrupole moments $Q_{\indrm{e}}/Q_{\indrm{g}}=-1.45(10)$ \cite{GMZ06}.} (MHz)  & $2.01$ \cite{BBS65}  $1.74$ \cite{RFI69}   & 0   \cite{CR57,SEB22}   & 15.5 - 24.4   & 0 & - \\
$\eta$    & 0 & 0 &  0.69 - 0 \cite{KRZ07,OT14} & 0 & -\\
\hline        
density (g/cm$^3$)   & 2.985 & 4.28 & 3.86   & 2.57 & 3.64\\
\hline        
Photoelectric absorption length $L_{\indrm{e}}$ ($\mu$m)   & 60  & 54.5 & 69.7  & 146 & 133  \\
\hline        
Sc number dens. $N_{\ind{0}}\! \times\!\! 10^{22}$ (1/cm$^3$)  &  3.98 & 4.37 & 3.18  & 1.54 & 0.88 \\
\hline        
Optimized optical thickness parameter     & 2.27 & 2.26 & 2.11  & 2.14 & 1.11 \\
$\xi^{*}=\sigma_{\indrm{R}} N_{\ind{0}} L_{\indrm{e}}/2$  ($f_{\indrm{LM}}=1$)  &  &  &   & & \\
\hline
Crystal thickness $L$ ($\mu$m)     & 120 & 110 & 140  & -- & 450 \\
\hline        
Optical thickness parameter $\xi=\sigma_{\indrm{R}} N_{\ind{0}} L/4$     & 2.3 & 2.3 & 2.1  & - & 1.9 \\
\hline
Thermal conductivity @300~K (W/m~K)   & 15.8  & 51-56 \cite{AZC20} & 17.3   & 9.6 \cite{KBA19} & 4 \cite{MMS23}\\
 \hline \hline        
\end{tabular}
\caption{Crystal targets considered for the \sca\ NFS experiment,
  their properties, and parameters.  ScF$_3$ single
    crystals were not available and therefore were not used in the
    experiment.}
\label{tab2}
\end{table*}

\noindent
{\bf Decay time of incoherent \sca\ fluorescence.}
To determine the lifetime from the data, we chose a suitable analysis
range of detection times after the x-ray excitation, and binned the
data in time using a variable number of bins. For each number of bins,
further variable shifts of the binning grid along the time axis were
considered. Subsequently, an exponential decay with rate
$\gamma=1/\taux$ was fitted to the binned data, taking into account
the statistical error in the measured photon number for low number of
counts. Note that $\gamma$ is a more reliable fit parameter than
$\taux$, since $\taux$ may become large and may even diverge in case
the fitted decay rate approaches zero. We found that due to the
statistical fluctuations of the measured data, the fit result depends
on the chosen analysis parameters, and they also vary within the
expected statistical fluctuations in between the two
detectors. Therefore, we combined the data of both detectors to
improve the statistics, and repeated the analysis many times, with
start time of the analysis chosen in the interval [30,40] ms, end time
of the analysis region chosen in the interval [88, 90] ms, number of
bins between 40 and 100, and including 10 shifts of the binning grid
within the range of one bin width.  { The fitting
  range was chosen to exclude time intervals affected by the motion of
  the mechanical shutter, which contain instrumental artifacts
  unrelated to the nuclear decay.}  We then created a histogram of the
fit results, which could be well-fitted using a Gaussian distribution
of decay rates. From the mean and the standard deviation of this fit,
we determine the lifetime $\taux$ and its error range reported in the
manuscript.

We note that the residual population of excited nuclei
  from earlier XFEL macropulses (memory effect) was taken into account in our
  analysis.  Importantly, in the case of a purely exponential decay
  law, the memory effect does not distort the exponential shape of the
  time distribution or bias the extraction of the decay constant. Each
  pulse contributes an independent exponentially decaying population,
  and the superposition of such contributions retains an overall
  exponential form with the same decay time. Thus, the extracted
  lifetime remains valid, even in the presence of pulse-to-pulse
  accumulation.

\noindent
{\bf Partial $K$-shell internal conversion coefficient
$\alpha_{\ind{K}}$} quantifies the probability of internal
conversion (electron emission) from the $K$-shell relative to $\gamma$-ray emission in a nuclear decay. It is the largest component of the total internal conversion coefficient $\alpha = \alpha_{\ind{K}} + \alpha_{\ind{L}} + \dots$, which includes all contributing electron shells.

In our experiment, $\alpha_{\ind{K}}$ is determined as
\begin{equation}
  \alpha_{\ind{K}}=\frac{R_{4}-2R_{B}}{R_{12}-2R_{B}}\,\,\, \frac{1}{\omega_{\ind{K}}} \,\,\, \frac{Y_{12}}{Y_{4}} = 390(60).
\label{eq103}
\end{equation}
Here, $R_{\ind{4}} = 328(6)$~ph/keV/10000~s is the Sc
$K_{\ind{\alpha,\beta}}$ fluorescence rate, and $R_{\ind{12}} = 7.3(0.9)$~~ph/keV/10000~~s is the 12.4-keV elastic fluorescence rate. Both are corrected for the background $2R_{\mathrm{B}} = 1.8$~counts/keV/10000~s.

Since $K$-shell conversion results in both $K_{\ind{\alpha,\beta}}$ fluorescence (detected) and Auger electron emission (undetected), the fluorescence yield must be included. For Sc, $\omega_{\ind{K}} = 0.19{0(2)}$ \cite{Krause79,HTS94}.

To account for the different attenuation of x-ray photons at energies $E$=4.5~keV and 12.4~keV in the 25-$\mu$m Sc foil, the relative fluorescence yield into both detectors is given by:
\begin{equation}
\label{eq105}
  \begin{split}
Y_{\ind{E}}=\frac{L_{\ind{1}}}{2L}\left[1-e^{-L/L_{\ind{1}}}\right]+\frac{L_{\ind{2}}}{2L}\left[1-e^{-L/L_{\ind{2}}}\right] e^{-L/L_{\ind{E}}},\\
\frac{1}{L_{\ind{1}}}=\frac{1}{L_{\ind{12}}}+\frac{1}{L_{\ind{E}}}, \hspace{1cm} \frac{1}{L_{\ind{2}}}=\frac{1}{L_{\ind{12}}}-\frac{1}{L_{\ind{E}}}, 
  \end{split}  
\end{equation}
where $L=25$~~$\mu$m is the foil thickness, and $L_{\indrm{e}}$ is the photoelectric absorption length: $L_{\ind{4}} = 27$~~$\mu$m for 4.1~~keV and $L_{\ind{12}} = 60$~~$\mu$m for 12.4~keV. This yields $Y_4 = 0.53$ and $Y_{12} = 0.67$.

Accounting for all corrections gives $\alpha_{\ind{K}} = 390(60)$, in agreement with the theoretical value $\alpha_{\ind{K}} = 363$ \cite{KBT08}, and consistent with the indirect result from \cite{SRK23}.\\

\noindent
{\bf \sca\ NFS crystal targets.}
Achieving minimal inhomogeneous linewidth broadening in \sca\ NFS targets and preventing further broadening during x-ray excitation are key challenges for successful NFS observation.

To address the former, we selected and prepared Sc-based crystalline (not polycrystalline) samples that met the following criteria:\
(a) highest crystallinity,\
(b) high chemical purity,\
(c) local cubic symmetry at Sc sites to nullify the electric field gradient (EFG) $V_{\ind{zz}}$ and eliminate \sca\ quadrupole hyperfine (HF) interactions,\
(d) nonmagnetic to suppress magnetic HF interactions,\
(e) minimal non-Sc atoms—preferably low-$Z$—to maximize photoelectric absorption length $L_{\indrm{e}}$ and NFS signal,\
(f) optimal thickness $L$ to maximize NFS signal, expected from Eq.~\eqref{eq001} at $L^{} = 2L_{\indrm{e}}$ \cite{SS90}, corresponding to optical thickness $\xi^{} = L_{\indrm{e}} / 2L_{\indrm{r}}$,\
(g) high thermal conductivity for efficient heat dissipation.

Cooling the targets to $\lesssim 30$~K was also important to avoid
the resonance broadening due to second-order Doppler effect, see
discussion in \cite{SRK23}. 

The following crystals were used in the experiment: Sc\footnote{A bulk
  Sc metal crystal was obtained from Ames National Laboratory, USA. It
  was cut into 250-$\mu$m thick plates in the (0001) and (0100)
  orientations using electrical discharge machining, etched to an
  optimal thickness of 120~$\mu$m in nitric acid, and annealed in
  vacuum at 800~°C for 24 hours.}, ScN \cite{AZC20}, \sco\
\cite{PBP02}, and \sam\footnote{\sam\ crystal plates of 0.45 mm
  thickness were obtained from Fukuda Crystal Laboratory Co., Ltd.,
  Sendai, Japan}, with selected properties listed in Table~\ref{tab2}.

None of the crystals fully satisfied all target criteria. ScN would be
the best choice under ideal conditions, but was
non-stoichiometric—ScN$_{0.8}$ with a 20\% nitrogen deficiency \cite{AZH22}.

\sam\ crystals exhibited the highest crystalline quality \cite{ISF21}
but had low thermal conductivity and a non-zero EFG.  X-ray Bragg
diffraction topography studies conducted on \sam\ crystals before and
after the experiment showed no evidence of radiation damage, within
the sensitivity limits of the topographic technique.

\noindent
{\bf Isomer excitation in  the NFS targets.}  
Even in the absence of a clear NFS signal,
the excitation of the isomer in the crystal samples is confirmed by two independent observations:\\
(1) the simultaneous detection of incoherent resonant \sca\ fluorescence (see Fig.~\ref{fig2}), and\\
(2) the direct detection of the prompt 12.4-keV photons in the forward
direction (see Fig.~\ref{fig4}, inset and main panel).

The prompt peak matches the expected energy and appears significantly
above the background level, despite attenuation from the tungsten
shutter blade. While delayed coherent re-emission is not observed
beyond 2~ms, the combination of the incoherent fluorescence and the
prompt forward signal confirms that excitation occurred.

All critical components of the NFS detection system—including detector
background, timing resolution, and shutter synchronization—were
individually characterized. The performance of the mechanical shutter,
which defines the accessible time window for observing delayed
signals, was independently verified using a laser beam. These tests
confirmed a long-term reliable opening duration of $<1$ ms (10–90\%
level), with low timing jitter of $\lesssim1$ ms, ensuring that any
coherent signal persisting beyond a 2 ms delay would have been
observable.

\noindent {\bf Minimizing radiation damage of the NFS targets.}  This
requirement was addressed by cooling the crystal targets to 20~K and
minimizing photon flux while maintaining high spectral flux via x-ray
monochromatization.  However, the 15-meV-bandwidth diamond channel-cut
monochromators designed for this purpose \cite{STB21,TPD22} failed to
function properly during the experiment. As a result, the direct XFEL
beam was used, likely contributing to the absence of observable \sca\
NFS.  The further reduction of the spectral bandwidth and total
radiation power on the NFS targets is foreseen in subsequent
experiments.

\end{document}